\documentclass[floatfix, preprint, showpacs, showkeys, preprintnumbers, nofootinbib, superscriptaddress]{revtex4-1}
\usepackage[utf8]{inputenc}
\usepackage[sort&compress]{natbib}
\usepackage{ulem}
\usepackage{bm}
\usepackage{times}
\usepackage{amssymb,amsbsy,amsmath,amsfonts}
\usepackage{graphicx}
\usepackage{epstopdf}
\usepackage{float}
\usepackage{color}
\usepackage{morefloats}
\usepackage{rotating}
\usepackage{srcltx}
\usepackage{slashed}
\usepackage{subfigure}
\usepackage{multirow}
\usepackage{verbatim}
\usepackage{hyperref}
\usepackage{overpic}
\usepackage{tabularx}

\begin{document}

\title{Triply charmed dibaryons in the  one boson exchange model}

\author{Ya-Wen Pan}
\affiliation{School of Physics, Beihang University, Beijing 100191, China}

\author{Ming-Zhu Liu}
\affiliation{School of Physics, Beihang University, Beijing 100191, China}

\author{Li-Sheng Geng}\email{lisheng.geng@buaa.edu.cn}
\affiliation{School of Physics, Beihang University, Beijing 100191, China}
\affiliation{
Beijing Key Laboratory of Advanced Nuclear Materials and Physics,
Beihang University, Beijing 100191, China}
\affiliation{School of Physics and Microelectronics, Zhengzhou University, Zhengzhou, Henan 450001, China}
\affiliation{Beijing Advanced Innovation Center for Big Data-Based Precision Medicine, School of Medicine and Engineering, Beihang University, Beijing, 100191}

\date{\today}
\begin{abstract}
  The pentaquark states, $P_{c}(4312)$, $P_{c}(4440)$ and $P_{c}(4450)$, can
    be nicely arranged into a multiplet of seven molecules of $\bar{D}^{(\ast)}\Sigma_{c}^{(\ast)}$  dictated
     by heavy quark spin symmetry, while the $\Xi_{cc}^{(\ast)}\Sigma_{c}^{(\ast)}$ system can be related to the $\bar{D}^{(\ast)}\Sigma_{c}^{(\ast)}$ system  via heavy antiquark diquark symmetry. In this work we employ  the one boson exchange model to study the interactions between $\Xi_{cc}^{(\ast)}$ and $\Sigma_{c}^{(\ast)}$ with constraints from the pentaquark system. We show that a  multiplet of ten triply charmed dibaryons emerge naturally in the isospin-1/2 sector, while only three appear in the isospin-3/2 sector.
      In addition, we study their  heavy quark flavor partners.
      Ten triply bottom diybaryons are found in the isospin-1/2 sector, while only nine are likely in the isospin-3/2 sector. Furthermore, the predicted mass splitting between
      the $0^{+}$ $\Xi_{cc}\Sigma_{c}$ state and its $1^{+}$ counterpart is found to be consistent with the correlation implied by heavy antiquark diquark symmetry recently pointed out in Pan $et$ $al$..
\end{abstract}


\maketitle

\section{Introduction}

  Since the Belle collaboration discovered $X(3872)$ in 2003~\cite{Choi:2003ue} and the BESIII collaboration discovered $Z_{c}(3900)$~\cite{Ablikim:2013mio} and $Z_{c}(4020)$~\cite{Ablikim:2013wzq},  many studies have
  been performed  to explore the existence of multiplets of hadronic molecules in the meson-meson sector. For instance, assuming
  that $X(3872)$ is a $1^{++}$ $\bar{D}D^{\ast}$ state,  a $2^{++}$ $\bar{D}^*D^*$ state has been predicted~\cite{Nieves:2012tt,Guo:2013sya} . With further assumptions on the nature of some other states, such as $X(3915)$~\cite{Nieves:2012tt} or $Z_b(10610)$~\cite{Guo:2013sya} or the adoption of more model-dependent approaches, such as the one-boson exchange (OBE) model~\cite{Liu:2019stu}, more states can be predicted.  Nevertheless,
  whether or not  a complete multiplet of molecules exists remains unsettled in the meson-meson sector, partly
  due to the fact that the weak attraction between $\bar{D}$ and $D^*$ deduced from  the small binding  of $X(3872)$  could easily
  disappear or even turn repulsive away from the strict heavy quark symmetry (HQS) limit,
  which might explain the continued absence of its spin-2 partner, $X(4012)$.

 In 2019 the LHCb collaboration surprised the hadron physics community by reporting the observation of
  three pentaquark states,
 $P_c(4312)$, $P_c(4440)$ and $P_c(4457)$~\cite{Aaij:2019vzc}, updating their 2015 study~\cite{Aaij:2015tga}. Given the fact that $P_c(4312)$ is located $9.8\,{\rm MeV}$ below the $\bar{D} \Sigma_c$
  threshold, and $P_c(4440)$ and $P_c(4457)$ are $21.8$ and
  $4.8\,{\rm MeV}$ below the $\bar{D}^* \Sigma_c$ one, they provide
  the most robust candidates so far for hadronic molecules.~\footnote{It should be noted
  that the existence of $\bar{D}^{(*)}\Sigma_c^{(*)}$ molecules had been predicted before the first LHCb results~\cite{Wu:2010jy,Wu:2010vk,Xiao:2013yca,Karliner:2015ina,Wang:2011rga,Yang:2011wz} and although at present
   the molecular interpretation is the most favored one, there exist other explanations, e.g.,
hadro-charmonium~\cite{Eides:2019tgv}, compact pentaquark states~\cite{Ali:2019npk,Wang:2019got,Cheng:2019obk,Weng:2019ynv,Zhu:2019iwm,Pimikov:2019dyr}, or
virtual states~\cite{Fernandez-Ramirez:2019koa}. See Refs.~\cite{Liu:2019zoy,Brambilla:2019esw,Guo:2019twa} for some latest reviews.}
Indeed, a large number of theoretical  works  have been performed within the molecular picture, focusing on various aspects ranging from mass spectroscopy~\cite{Liu:2019tjn,Chen:2019bip,Chen:2019asm,He:2019ify,Xiao:2019aya,Meng:2019ilv,Yamaguchi:2019seo,Liu:2019zvb,Burns:2019iih,Valderrama:2019chc,Wang:2019ato,Chen:2020pac,Xu:2020gjl,Zhang:2020erj} and decay widthes~\cite{Xiao:2019mst,Sakai:2019qph,Lin:2019qiv,Wang:2019spc,He:2019rva,Guo:2019kdc} to production mechanisms~\cite{Wang:2019krd,Wu:2019rog,Wang:2019dsi,Yang:2020eye,Xie:2020niw}.
 In our previous work we adopted the contact range effective field theory (EFT) to describe the three pentaquark states in the hadronic molecular picture, and we showed the emergence of a complete multiplet of hadronic molecules in the meson-baryon system~\cite{Liu:2019tjn}, dictated by heavy quark spin symmetry (HQSS)~\cite{Isgur:1989vq,Isgur:1989ed}, which has later been  corroborated by many studies~\cite{Liu:2019tjn,Xiao:2019aya,Yamaguchi:2019seo,Liu:2019zvb,Valderrama:2019chc,Du:2019pij}.
  %
  %

   Heavy antiquark diquark symmetry (HADS) dictates that a heavy antiquark behaves the same
   as a heavy diquark from the perspective of the strong interaction in the limit of heavy quark masses~\cite{Savage:1990di,Fleming:2005pd}. Therefore, the strong interaction experienced by a heavy anticharm meson is the same as that of a doubly charmed baryon. As a result,   the interaction between $\bar{D}^{(\ast)}$ and $\Sigma_{c}^{(\ast)}$ is the same as that between $\Xi_{cc}^{(\ast)}$ and $\Sigma_{c}^{(\ast)}$, as shown in Fig.~1, up to the breaking of HADS.   In Ref.~\cite{Pan:2019skd}, HADS was utilized  in the contact range EFT to
explore the likely existence of $\Xi_{cc}^{(\ast)}\Sigma_{c}^{(\ast)}$ dibaryons implied by the existence of the pentaquark states, and a complete multiplet of ten dibaryons emerges even when taking into account the breaking of HADS at the level of 25\%.
Another interesting finding of Ref.~\cite{Pan:2019skd} is the existence of a strong correlation between
  the mass splitting of any doublet in the dibaryon system and the spin assignments of
    $P_c(4440)$ and $P_c(4457)$. Though a lattice QCD study of the pentaquark system is difficult~\cite{Sugiura:2019pye,Skerbis:2018lew}, the study of
  the dibaryon systems seems to be relatively easier and has recently been performed in Ref.~\cite{Junnarkar:2019equ}. However, unfortunately, only
  the deuteron-like $1^+$ states were studied. If the spectrum of the $0^+$ states can be obtained, one could determine the spins of 
  $P_c(4440)$ and $P_c(4457)$ model independently using the correlation dictated by HADS, as first
  observed in Ref.~\cite{Pan:2019skd}.

  %

\begin{figure}[!h]
 \center{\includegraphics[width=12cm]  {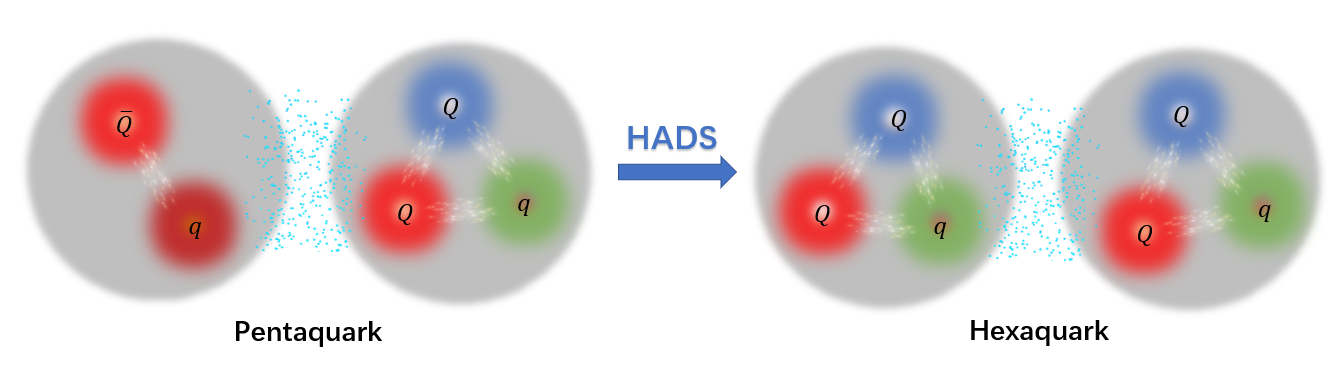}}
 \caption{\label{cha}From hidden charm pentaquark states to triply charmed hexaquark dibaryons through HADS}
 \end{figure}

In this work we adopt the OBE model to study the interactions between $\Xi_{cc}^{(\ast)}$ and $\Sigma_{c}^{(\ast)}$.
The OBE model is one of the most widely used theoretical tools in studying heavy hadron molecules, see, e.g., Refs.~\cite{Machleidt:1987hj,Chen:2016qju}. According to this model, the potential between two hadrons is generated by the exchange of a series of light mesons, such as the $\pi$, $\sigma$, $\rho$, and $\omega$. Though physically intuitive and numerically clean,
the OBE model suffers from the need for regulators (form factors and cutoffs) to mimic the finite sizes of hadrons involved.
 There is no a priori information for which cutoff to use~\footnote{We implicitly assume that
different regulators should yield the same physical results, if the theory is properly normalized, and focus on the
impact of the value of the cutoff in a cutoff regularization scheme.}, though according to the size of typical hadrons, a value of 0.5 to 1 GeV is preferred~\cite{Yamaguchi:2011xb,Liu:2008fh}. In the nucleon-nucleon system, because of the large nucleon-nucleon scattering dataset, the form factors and related cutoffs can be determined by fitting to the large amount of data~\cite{Machleidt:2000ge}. In the heavy sector, however data are scarce, and as a result, in most cases the cutoff has to be determined by
fitting to the same data that one wants to describe, therefore giving the impression that the model has no predictive power.
In this sense, the OBE model should better be  used to estimate the interaction between two hadrons by varying the cutoffs within a reasonable range. On the other hand, in a few cases where there is  a clean bound-state candidate of two hadrons, one can fix the cutoff by reproducing such a state first and then make predictions for other related systems using symmetry arguments.  This is the case for the present study.
  In this work, following Ref.~\cite{Liu:2019zvb} and assuming that $P_{c}(4312)$ is a bound state of $\bar{D}\Sigma_{c}$, we can fix the cutoff, i.e.,  $\Lambda=1.119$ GeV. With this cutoff and the OBE potential, we study  the $\Xi_{cc}^{(\ast)}\Sigma_{c}^{(\ast)}$  system.

  The manuscript is structured as follows: In Sec. ~\ref{sec:obe}, we present the details of the OBE
model as applied to the baryon-baryon system containing heavy quarks $c$ and $b$.
In Sec. ~\ref{sec:pre}, we determine the cutoff in the OBE model by
 reproducing $P_{c}(4312)$ as a hadronic molecule, and from this
we predict the full spectrum of triply heavy dibaryon molecules.
Finally we present the conclusions in Sec. \ref{sec:conclusions}.
\section{Formulism}
\label{sec:obe}

In this section, we derive the OBE potentials that we use in this work.
The OBE interaction of two heavy hadrons is generated by the exchange of $\pi$, $\sigma$, $\rho$, and $\omega$.
Among them,  the vector mesons, $\rho$ and $\omega$, provide the short-range interaction,  the
scalar meson $\sigma$ provides the middle-range  interaction, and the $\pi$ meson provides the long-range interaction.
Given the exploratory nature of the present work, the contributions of other mesons are neglected~\cite{Liu:2019zvb,Meng:2017fwb}.

\subsection{Interaction Lagrangian}

The Lagrangians describing the interaction between a
doubly heavy charmed baryon  and a light meson, $\rho$, $\omega$, $\sigma$, and $\pi$,
 can be written as~\cite{Liu:2019stu}
\begin{eqnarray}
    \mathcal{L}_{T_{cc}T_{cc}\pi} &=& \frac{ig_{1}}{\sqrt{2} f_{\pi}}
{\vec{T}_{cc}}^{\dagger}  \cdot ( \vec{\tau} \cdot \vec{\pi} \vec{ \nabla} \times
 \vec{T}_{cc}) \, , \label{eq:L-pi} \\
  \mathcal{L}_{T_{cc}T_{cc}\sigma} &=& g_{\sigma1}
    {\vec{T}_{cc}}^{\dagger} \sigma\cdot \vec{T}_{cc}\, , \label{eq:L-sigma} \\
  \mathcal{L}_{T_{cc}T_{cc} \rho} &=& g_{\rho1}
    {\vec{T}_{cc}}^{\dagger} (\vec{\tau} \cdot \vec{\rho}^{0}) \cdot \vec{T}_{cc}
  \nonumber \\
  &-& \frac{f_{\rho 1}}{4 M_{1}}\,
    {\vec{T}_{cci}}^{\dagger}  \vec{\tau} \cdot \left( \partial^i \vec{\rho}^j
    - \partial^j \vec{\rho}^i \right)\vec{T}_{ccj}  \, , \label{eq:L-rho}\\
    \mathcal{L}_{T_{cc}T_{cc}\omega} &=& g_{\omega1}
    {\vec{T}_{cc}}^{\dagger} {\omega}^{0}\cdot  \vec{T}_{cc}
  \nonumber \\
  &-& \frac{f_{\omega1}}{4 M_{1}}\,
      {\vec{T}_{cci}}^{\dagger}  \, \left( \partial^i {\omega}^j
    - \partial^j {\omega}^i \right) \vec{T}_{ccj} \, , \label{eq:L-omega}
\end{eqnarray}
where $\vec{T}_{cc}=(\frac{1}{\sqrt{3}}\Xi_{cc}\vec{\sigma}+\vec{\Xi}_{cc}^{\ast})$ is a superfield of $\Xi_{cc}$ and $\vec{\Xi}_{cc}^{\ast}$  constrained  by HQSS. With HADS, these Lagrangians can also be derived from
 the Lagrangians describing the interaction between a heavy meson and a light meson \cite{Liu:2018zzu}.
 The  $g_{1}$ and $g_{\sigma}$ are the couplings to the $\pi$ and $\sigma$ meson, respectively, while the $g_{v}$ and $f_{v}$ with $V=\rho, \omega$ are the electro- and magnetic-type couplings to the vector mesons, and $M_{1}$ is a mass scale rendering $f_{v}$ dimensionless.

For the Lagrangians describing the interaction between a singly charmed baryon  and  a light meson, we have~\cite{Liu:2019zvb}
\begin{eqnarray}
    \mathcal{L}_{S_{c}S_{c}\pi} &=& \frac{ig_{2}}{\sqrt{2} f_{\pi}}
{\vec{S}_{c}}^{\dagger} \cdot ( \vec{T} \cdot \vec{\pi}\vec{\nabla}
 \times \vec{S}_{c}) \, , \label{eq:L-pi} \\
  \mathcal{L}_{S_{c}S_{c}\sigma} &=& g_{\sigma2}
    {\vec{S}_{c}}^{\dagger} \sigma\cdot  \vec{S}_{c} \, , \label{eq:L-sigma} \\
  \mathcal{L}_{S_{c}S_{c} \rho} &=& g_{\rho2}
    {\vec{S}_{c}}^{\dagger} (\vec{T} \cdot \vec{\rho}^{0}) \cdot \vec{S}_{c}
  \nonumber \\
  &-& \frac{f_{\rho 2}}{4 M}\,
    {\vec{S}_{ci}}^{\dagger}  \vec{T} \cdot \left( \partial^i \vec{\rho}^j
    - \partial^j \vec{\rho}^i \right)\vec{S}_{cj}  \, , \label{eq:L-rho} \\
    \mathcal{L}_{S_{c}S_{c}\omega} &=& g_{\omega2}
    {\vec{S}_{c}}^{\dagger} {\omega}^{0}\cdot \vec{S}_{c}
  \nonumber \\
  &-& \frac{f_{\omega2}}{4 M}\,
    {\vec{S}_{ci}}^{\dagger} \, \left( \partial^i {\omega}^j
    - \partial^j {\omega}^i \right) \vec{S}_{cj}  \, , \label{eq:L-omega}
\end{eqnarray}
where $\vec{S}_{c}=(\frac{1}{\sqrt{3}}\Sigma_{c}\vec{\sigma}+\vec{\Sigma}_{c}^{\ast} )$ denote the superfield of $\Sigma_{c}$ and $\Sigma_{c}^{\ast}$ dictated by HQSS, and $g_{2}$, $g_{\sigma2}$, $g_{v2}$ and $f_{v2}$ are
the corresponding coupling constants.

\subsection{The OBE potentials}

With the Lagrangians for the doubly charmed and singly charmed baryons,  the OBE potentials can be easily derived as follows,
\begin{eqnarray}
  V = \zeta \, V_{\pi} + V_{\sigma} + V_{\rho} + \zeta \, V_{\omega} \, ,
\end{eqnarray}
where $\zeta=\pm 1$, for which our convention is
\begin{eqnarray}
  \zeta = +1 && \quad\mbox{for $T_{cc}S_{c}$} \, , \\
  \zeta = -1 && \quad\mbox{for $\bar{T}_{cc} S_{c}$} \, .
\end{eqnarray}
and $V_{\pi,\sigma,\rho,\omega}$ are the OBE poentials in momentum space, whose explicit form can be found in, e.g., Refs.~\cite{Liu:2019stu,Liu:2019zvb}.

To take into account the finite sizes of hadrons, we introduce a monopolar form factor for each vertex, and then the
resulting potentials in momentum space can be written as
\begin{eqnarray}
  V_M(\vec{q}, \Lambda) = V_M(\vec{q})\,F_{1}(\vec{q},\Lambda_{1})F_{2}(\vec{q},\Lambda_{2})\,
  \, ,
\end{eqnarray}
where the monopolar form factor is
\begin{eqnarray}
  F(q, \Lambda)=
 \frac{\Lambda^{2}-m^2}{\Lambda^{2}-{q}^2}
  \label{Eq:FF} \, .
\end{eqnarray}
For the sake of simplicity, we adopt the same cutoff for  both vertices appearing in the one-boson exchange.
With the consideration of form factors, the coordinate-space potentials can be easily obtained by Fourier transforming the momentum-space potentials. For details, please refer to Refs.~\cite{Liu:2019stu,Liu:2019zvb}.

Note that in the present work, following Ref.~\cite{Liu:2019zvb},
 we remove the Dirac delta potential, $\delta(\vec{r})$, in the spin-spin component of the OBE potential.
 The delta potential, as can be seen
 in Fig.~2 of Ref.~\cite{Liu:2019zvb}, distorts the long-range one pion exchange potential at a distance of about 1 fm, thus
 obscuring the physics of the pion exchange, which given its long range, is expected to be model independent and should not be distorted. In Ref.~\cite{Liu:2019zvb}, it was shown that
 only by removing this delta potential can one achieve a simultaneous description of the three pentaquark states with
 a single cutoff fixed by reproducing either of the three states. Otherwise, one will be forced
 to adopt different cutoffs for different pentaquark states.
 We note that the RCNP group also follows such a procedure in their use of the OBE model~\cite{Yamaguchi:2011qw,Ohkoda:2011vj,Yamaguchi:2019vea}.

\subsection{Coupling constants }
\begin{table}[!h]
\centering
\caption{Couplings of the doubly and singly charmed baryons to the light mesons. The magnetic coupling of the $\rho$ and $\omega$ mesons is defined as $f_{v}=\kappa_{v}g_{v}$, and
$M$ refers to the mass scale (in MeV) involved in the magnetic-type couplings.
}
\label{tab:couplings2}
\begin{tabular}{cc|ccc}
  \hline \hline
  Coupling  & Value for $\Xi_{cc}$/$\Xi_{cc}^*$   &  Coupling  & Value for $\Sigma_{c}$/$\Sigma_{c}^*$\\
  \hline
  $g_1$ & -0.2          &    $g_2$ & 0.84 \\
  $g_{\sigma 1}$ & 3.4 &    $g_{\sigma 2}$ & 6.8 \\
  $g_{\rho 1}$ & 2.6 &   $g_{\rho 2}$ & 5.8 \\
  $g_{\omega 1}$ & 2.6&  $g_{\omega 2}$ & 5.8 \\
  $\kappa_{\rho 1}$ & -0.8 &  $\kappa_{\rho 2}$ & 1.7 \\
  $\kappa_{\omega 1}$ &-0.8 &  $\kappa_{\omega 2}$ &1.7\\
  $M_1$ & 940 &  $M_1$ & 940 \\
  \hline \hline
\end{tabular}
\end{table}

The OBE potentials are determined by the couplings of the exchanged bosons to the
heavy baryons and mesons. In this section, we explain how they are fixed and estimate how uncertain they are.
The coupling of the $\pi$  to the doubly charmed baryon can be  derived from either the quark model or HADS. Both approaches yield a similar value: $g_{1}=-0.25$~\cite{Liu:2018bkx} from the quark model and $g_{1}=-0.2$~\cite{Liu:2018euh} from HADS, and in this work we adopt $-0.2$. The coupling of the $\pi$ to $\Sigma_{c}$ and $\Sigma_{c}^{\ast}$,
$g_{2}$, was extracted to be $0.84$ in lattice QCD~\cite{Detmold:2012ge}, which is smaller than the prediction of the quark model~\cite{Liu:2011xc}.

For the couplings to the $\sigma$ meson, we estimate them using the
 quark model.  From the nucleon and $\sigma$-meson coupling of the
 liner sigma model, $g_{\sigma NN}=10.2$, the corresponding couplings  are determined to  be $g_{\sigma1}=\frac{1}{3}g_{\sigma NN}=3.4$ and $g_{\sigma2}=\frac{2}{3}g_{\sigma NN}=6.8$~\cite{GellMann:1960np}.

The couplings of the light vector mesons include those of both
electric($g_{v}$) and magnetic($f_{v}$)types, which are related via $f_{v}=\kappa_{v}g_{v}$.
For the $\rho$ and $\omega$ couplings, from SU(3)-symmetry and the OZI rule, we obtain $g_{\omega}=g_{\rho}$ and $f_{\omega}=f_{\rho}$.
The electric-type coupling of the doubly charmed baryon  to the vector mesons is estimated to be $g_{v1}=2.6$~\cite{Casalbuoni:1992dx}, and the corresponding  coupling to the singly charmed baryon
is determined to be $g_{v2}=5.8$~\cite{Liu:2011xc}.  The magnetic-type coupling of the $\Sigma_{c}^{(\ast)}$ baryon is estimated to be $\kappa_{v2}=1.7$~\cite{Can:2013tna}, and the corresponding coupling of the $\Xi_{cc}^{(\ast)}$ baryon is $\kappa_{1v}=-0.8$. All the couplings are given in Table \ref{tab:couplings2} for easy reference.   In addition, the $\Sigma_{c}^{(\ast)}$, $\Sigma_{b}^{(\ast)}$,  and $\Xi_{cc}$ masses are from Ref.~(\cite{Tanabashi:2018oca}), and the $\Xi_{cc}^{\ast}$ and $\Xi_{bb}^{(\ast)}$ masses are taken from lattice QCD simulations~\cite{Lewis:2008fu}.

\subsection{Wave functions and partial wave decomposition}

\begin{table*}[t]
\centering
\centering \caption{Relevant partial wave matrix elements for the $\Xi_{QQ}^{(\ast)}\Sigma_{Q}^{(\ast)}$ system
with $Q=c$ or $b$.} \label{tab:spin1123}
\begin{tabular}{c|c|c|ccccccc}
\hline\hline
State & $J^{P}$  &   Partial wave    & $\langle \vec{a}_{1}\cdot \vec{a}_{2}\rangle$& $S_{12}(\vec{a}_{1}, \vec{a}_{2}, \vec{r})$ \\ \hline \hline
 $\Xi_{QQ}\Sigma_{Q}$  &  $J=0$   &   $^1S_{0}$& -3  & 0
\\ \hline
 $\Xi_{QQ}\Sigma_{Q}$  & $J=1$ & $^3S_{1}$-$^3D_{1}$    & $
\left(\begin{matrix}
1 & 0 \\
0 & 1\\
\end{matrix}\right)$
 & $
\left(\begin{matrix}
0 & \sqrt{8} \\
\sqrt{8} & -2\\
\end{matrix}\right)$
\\ \hline
 $\Xi_{QQ}\Sigma_{Q}^{\ast}$ /$\Xi_{QQ}^{\ast}\Sigma_{Q}$  & $J=1$ & $^3S_{1}$-$^3D_{1}$-$^5D_{1}$  &
  $
\left(\begin{matrix}
-\frac{5}{2} & 0 & 0\\
0 & -\frac{5}{2}& 0\\
  0    &0          &\frac{3}{2}  \\
\end{matrix}\right)$  &  $
\left(\begin{matrix}
0 & -\frac{1}{\sqrt{2}} & -\frac{3}{\sqrt{2}}\\
-\frac{1}{\sqrt{2}}  & \frac{1}{2}& -\frac{3}{2}\\
 -\frac{3}{\sqrt{2}}    &-\frac{3}{2}        &-\frac{3}{2}  \\
\end{matrix}\right)$
\\ \hline
 $\Xi_{QQ}\Sigma_{Q}^{\ast}$ /$\Xi_{QQ}^{\ast}\Sigma_{Q}$  & $J=2$ & $^5S_{2}$-$^3D_{2}$-$^5D_{2}$  & $
\left(\begin{matrix}
\frac{3}{2} & 0 & 0\\
0 & -\frac{5}{2}& 0\\
  0    &0          &\frac{3}{2}  \\
\end{matrix}\right)$  &  $
\left(\begin{matrix}
0 & 3\sqrt{\frac{3}{10}} & 3\sqrt{\frac{7}{10}}\\
3\sqrt{\frac{3}{10}}   & -\frac{1}{2}& -\frac{3}{2}\sqrt{\frac{3}{7}}\\
3\sqrt{\frac{7}{10}}    &-\frac{3}{2}\sqrt{\frac{3}{7}}      &\frac{9}{14}  \\
\end{matrix}\right)$
\\ \hline
 $\Xi_{QQ}^{\ast}\Sigma_{Q}^{\ast}$  & $J=0$ & $^1S_{0}$-$^5D_{0}$
& $
\left(\begin{matrix}
-\frac{15}{4} & 0 \\
0 & -\frac{3}{4}\\
\end{matrix}\right)$  &$
\left(\begin{matrix}
0 & -3 \\
-3 & -3\\
\end{matrix}\right)$   \\ \hline
 $\Xi_{QQ}^{\ast}\Sigma_{Q}^{\ast}$  & $J=1$ & $^3S_{1}$-$^3D_{1}$-$^5D_{1}$-$^7D_{1}$  &  $
\left(\begin{matrix}
-\frac{11}{4} & 0  &0 &0\\
0 & -\frac{11}{4} & 0  &0\\
0 & 0&  -\frac{3}{4} & 0  \\
0 & 0&  0&  \frac{9}{4} \\
\end{matrix}\right)$  &$
\left(\begin{matrix}
0 & \frac{17}{5\sqrt{2}}  &0 &-\frac{3\sqrt{7}}{5}\\
\frac{17}{5\sqrt{2}} & -\frac{17}{10} & 0  &\frac{3}{5}\sqrt{\frac{2}{7}}\\
0 & 0&  -\frac{3}{2} & 0  \\
-\frac{3\sqrt{7}}{5} & \frac{3}{5}\sqrt{\frac{2}{7}}&  0 &  -\frac{108}{35} \\
\end{matrix}\right)$   \\ \hline
 $\Xi_{QQ}^{\ast}\Sigma_{Q}^{\ast}$  & $J=2$ & $^5S_{2}$-$^1D_{2}$-$^3D_{2}$-$^5D_{2}$-$^7D_{2}$   & $
\left(\begin{matrix}
-\frac{3}{4} & 0  &0 &0 &0\\
0 & -\frac{15}{4} & 0  &0  &0\\
0 & 0&  -\frac{11}{4} & 0 &0 \\
0 & 0&  0&  -\frac{3}{4} &0\\
0 & 0&  0&  0& \frac{9}{4} \\
\end{matrix}\right)$  &$
\left(\begin{matrix}
0 & -\frac{3}{\sqrt{5}}  &0 &3\sqrt{\frac{7}{10}} &0\\
-\frac{3}{\sqrt{5}}  & 0 & 0  & 3\sqrt{\frac{2}{7}}  &0\\
0 & 0&  \frac{17}{10} & 0 &\frac{6}{5}\sqrt{\frac{3}{7}} \\
3\sqrt{\frac{7}{10}} & 3\sqrt{\frac{2}{7}}&  0&  \frac{9}{14} &0\\
0 & 0&   \frac{6}{5}\sqrt{\frac{3}{7}}& 0 & -\frac{27}{35} \\
\end{matrix}\right)$
\\ \hline
 $\Xi_{QQ}^{\ast}\Sigma_{Q}^{\ast}$  & $J=3$ & $^7S_{3}$-$^3D_{3}$-$^5D_{3}$-$^7D_{3}$  & $
\left(\begin{matrix}
\frac{9}{4} & 0  &0 &0\\
0 & -\frac{11}{4} & 0  &0\\
0 & 0&  -\frac{3}{4} & 0  \\
0 & 0&  0&  \frac{9}{4} \\
\end{matrix}\right)$  &$
\left(\begin{matrix}
0 & -\frac{3\sqrt{3}}{5}  &0 &\frac{5\sqrt{3}}{9}\\
-\frac{3\sqrt{3}}{5}  & -\frac{17}{35} & 0  &\frac{16}{35}\\
0 & 0&  \frac{12}{7} & 0  \\
\frac{5\sqrt{3}}{9} & \frac{16}{35}&  0 &  \frac{99}{70} \\
\end{matrix}\right)$
\\ \hline\hline
\end{tabular}
\end{table*}

The generic wave function of a baryon-baryon system reads
\begin{eqnarray}
  | \Psi \rangle = \Psi_{J M}(\vec{r}) | I M_I \rangle\, ,
\end{eqnarray}
where $| I M_I \rangle$ denotes the isospin wave function and   $\Psi_{J M}(\vec{r})$ denotes the spin and spatial wave function.
The dynamics in isospin space is embodied in the isospin factor $\tau\cdot T$.  In this work the total isospin is either $1/2$ or $3/2$ for the  $\Xi_{cc}^{(\ast)}\Sigma_{c}^{(\ast)}$ system, and the corresponding isospin factor are $-2$ and 1, respectively.

The spin wave function can be written as a sum over partial wave functions, which can be written as (in
the spectroscopic notation)
\begin{eqnarray}
  |{}^{2S+1}L_{J}\rangle &=& \sum_{M_{S},M_L}
  \langle L M_L S M_S | J M \rangle \, | S M_S \rangle \, Y_{L M_{L}}(\hat{r})
  \, , \nonumber \\
\end{eqnarray}
where $\langle L M_L S M_S | J M \rangle$ are the Clebsch-Gordan coefficients,
$| S M_S \rangle$ is the spin wavefunction, and $Y_{L M_L}(\hat{r})$ is the spherical harmonics.

In the partial wave decomposition of
the OBE potential, we encounter both spin-spin and tensor components
\begin{eqnarray}
  C_{12} &=& \vec{\sigma}_1 \cdot \vec{\sigma}_2 \, , \\
  S_{12} &=& 3 \vec{\sigma}_1 \cdot \hat{r}\,\vec{\sigma}_2 \cdot \hat{r} -
  \vec{\sigma}_1 \cdot \vec{\sigma}_2 \, ,
\end{eqnarray}
  In the present study, we
  consider both $S$ and $D$ waves.~\footnote{ However, the $D$-wave contributions are found to only account for a few percent of
 the binding energies.} The relevant matrix elements are listed
  in  Table~\ref{tab:spin1123}.

 We would like to note that in principle, some of the particle channels can mix: for instance,  for the $I(J^P)=1/2(1^+)$ states, $\Sigma_c^* \Xi_{cc}$ and $\Sigma_c\Xi_{cc}^*$ can mix with $\Sigma_c\Xi_{cc}$. We checked that our results in the single-channel case remain qualitatively the same
  when coupled channels effects are taken into account. Given the exploratory nature of the present study, we stick to the single-channel dominance assumption.

\section{Results and Discussion}
\label{sec:pre}

The main ambiguity of the OBE model is the value of the cutoff in the form factor,  which in principle is a free parameter in the range of 0.5$\sim$2 GeV. However, cutoffs in this range heavily  affect the resulting interaction.
 %
One way to make concrete predictions is to fix
the cutoff using a well-established molecular state as a reference, and then with this value, one predicts
interactions in other related systems and studies
the likely existence of molecular states. Such an approach has
been adopted in a number of studies. For instance, assuming that
 $X(3872)$ is a hadronic molecule of $\bar{D}D^{\ast}$,  the cutoff was determined to
$\Lambda=1.04$GeV~\cite{Liu:2019stu}.
In Ref~\cite{Liu:2019zvb}, the cutoff was determined to be $\Lambda=1.119$ GeV, assuming
 that  $P_{c}(4312)$ is a bound state of $\bar{D}\Sigma_{c}$.
%
 Thus, in this work we adopt $\Lambda=1.119$ GeV to study the triply charmed and bottom dibaryon systems.

Since we have used HADS in deriving the potentials of the $\Xi_{cc}^{(\ast)}\Sigma_{c}^{(\ast)}$ system, we should study
the breaking of the HADS, which is only exact in the infinite heavy quark mass limit.
The breaking of HADS can be estimated, as a rule of thumb,  by calculating $\Lambda_{QCD}/m_{Q}v$ with $v$ being the velocity of the heavy diquark pair~\cite{Savage:1990di}, which yields a value of $0.25-0.4$ for the charm system. In the present work, following Ref.\cite{Pan:2019skd}, we take 25\% as an educated guess. The change of our potentials induced by such a breaking can be taken into account by modifying  the OBE potentials in the following way:
\begin{eqnarray}
V=V_{OBE}(1+\delta_{HADS}),
\end{eqnarray}
where $V_{OBE}$ is the central value of the OBE potentials derived above and $\delta_{HADS}=0.25$ is the uncertainty induced by HADS.

With all the ingredients explained above we can solve the Scr$\ddot{o}$dinger equation in coordinate space, and study
 the spectrum of the $\Xi_{cc}^{(\ast)}\Sigma_{c}^{(\ast)}$ system. Those with $I=1/2$ are presented in Table \ref{tab:binding-hidden-dxicc189}.  Interestingly, we find ten  triply charmed dibaryons  with  binding energies of $10\sim40$ MeV.
In our previous study, we employed the contact range EFT constrained by HQSS and HADS to predict ten triply charmed dibaryons from the LHCb pentaquark states ~\cite{Pan:2019skd}. Our current results are consistent with
those of Scenario B of that study. In addition, the predicted binding energy of the  $1^+$ $\Xi_{cc}\Sigma_c$ state, $B=21.6^{+20.1}_{-14.6}$ MeV, is consistent with the lattice QCD result of Ref.~\cite{Junnarkar:2019equ}, which is $8\pm17$ MeV.~~\footnote{ As mentioned in Ref.~\cite{Junnarkar:2019equ}, lattice artifacts, such as finite volume effects, should be carefully addressed in a future work, particularly for the $\Sigma_c\Xi_{cc}$ dibaryon of spin-parity $1^+$. As a result, the quoted lattice QCD number should be taken with caution.}

\begin{table}[!h]
\centering
\caption{Binding energies of the triply charmed and bottomed  dibaryons with $I=1/2$, with
uncertainties originating from the breaking of HADS, and the corresponding masses, for which only central values are given.}
\label{tab:binding-hidden-dxicc189}
\begin{tabular}{ccc|cc|cccc}
\hline\hline
Molecule  & $I$ & $J^{P}$  & $B_{H_{ccc}}$ (MeV) & $M_{H_{ccc}}$ (MeV)   &   $B_{H_{bbb}}$ (MeV) & $M_{H_{bbb}}$ (MeV)    \\
\hline
$\Sigma_{Q}\Xi_{QQ}$ & $\frac{1}{2}$ & $0^+$ & $29.4^{+24.1}_{-18.4}$ &   6046   & $89.7^{+32.8}_{-30.1}$  & 15850\\
$\Sigma_{Q}\Xi_{QQ}$ & $\frac{1}{2}$ & $1^+$ & $21.6^{+20.1}_{-14.6}$ &   6053   & $76.1^{+29.0}_{-26.5}$  & 15864\\
\hline
$\Sigma_{Q}^{\ast}\Xi_{QQ}$ & $\frac{1}{2}$ & $1^+$ & $29.8^{+24.2}_{-18.6}$ &   6109  & $89.3^{+32.8}_{-30.1}$ &   15872 \\
$\Sigma_{Q}^{\ast}\Xi_{QQ}$ & $\frac{1}{2}$ & $2^+$ & $21.4^{+19.9}_{-14.5}$ &   6118  & $74.5^{+28.6}_{-26.0}$ &   15886  \\
\hline
$\Sigma_{Q}\Xi_{QQ}^{\ast}$ & $\frac{1}{2}$ & $1^+$ & $15.6^{+16.7}_{-11.5}$ &   6167  & $64.3^{+26.1}_{-23.3}$ &  15900    \\
$\Sigma_{Q}\Xi_{QQ}^{\ast}$ & $\frac{1}{2}$ & $2^+$ & $34.5^{+27.0}_{-21.0}$ &   6148  & $99.6^{+36.1}_{-33.1}$ &  15861    \\
\hline
$\Sigma_{Q}^{\ast}\Xi_{QQ}^{\ast}$ & $\frac{1}{2}$ & $0^+$ & $13.2^{+15.2}_{-10.2}$ &   6237  & $58.6^{+24.1}_{-21.7}$ &   15926 \\
$\Sigma_{Q}^{\ast}\Xi_{QQ}^{\ast}$ & $\frac{1}{2}$ & $1^+$ & $16.5^{+17.3}_{-12.0}$ &   6230  & $65.2^{+26.0}_{-23.6}$ &   15920\\
\hline
$\Sigma_{Q}^{\ast}\Xi_{QQ}^{\ast}$ & $\frac{1}{2}$ & $2^+$ & $24.7^{+21.8}_{-16.2}$ &   6222   & $80.7^{+30.7}_{-27.9}$ &   15904 \\
$\Sigma_{Q}^{\ast}\Xi_{QQ}^{\ast}$ & $\frac{1}{2}$ & $3^+$ & $35.7^{+27.0}_{-21.3}$ &   6211   & $98.5^{+35.3}_{-32.6}$ &   15886\\
  \hline\hline
\end{tabular}
\end{table}

The spins of $P_{c}(4440) $ and $P_{c}(4457)$ are not determined experimentally yet, while in Ref.~\cite{Pan:2019skd} we discovered that the mass splittings between any  doublets of the triply charmed dibaryons are correlated with that of $P_{c}(4440) $ and $P_{c}(4457)$. For instance, if the $0^{+}$ $\Xi_{cc}\Sigma_{c}$ state has a larger mass than its $1^{+}$ counterpart, then the spin of $P_c(4457)$ is $3/2$ and
that of $P_c(4440)$ is $1/2$. On the other hand, if the $0^{+}$ $\Xi_{cc}\Sigma_{c}$-state mass is smaller than $1^{+}$,  then the opposite assignment would be favored.
The present OBE results for the $0^{+}$ $\Xi_{cc}\Sigma_{c}$ and the $1^{+}$ $\Xi_{cc}\Sigma_{c}$ seem to prefer Scenario B, which indicates that  the spins of $P_{c}(4440)$ and $P_{c}(4457)$ should be  3/2 and 1/2, respectively.

\begin{table}[!h]
\centering
\caption{Binding energies of  the triply charmed and bottomed dibaryons  (if they exist) with $I=3/2$, with uncertainties originating
 from the breaking of HADS, and the corresponding masses, for which only central values are given.
 The $\dag$ indicates that the particular channel does not bind, and ? denotes the likely existence of
  a bound state.}
\label{tab:binding-hidden-dxicc190}
\begin{tabular}{ccc|cc|cccc}
\hline\hline
Molecule  & $I$ & $J^{P}$  & $B_{H_{ccc}}$ (MeV) & $M_{H_{ccc}}$ (MeV)  & $B_{H_{bbb}}$ (MeV) & $M_{H_{bbb}}$ (MeV)\\
\hline
$\Sigma_{Q}\Xi_{QQ}$ & $\frac{3}{2}$ & $0^+$ & $\dag$ &  $\dag$   & $1.8^{+2.9}_{-1.6}$&  15938    \\
$\Sigma_{Q}\Xi_{QQ}$ & $\frac{3}{2}$ & $1^+$ & $\dag$ &  $\dag$   & $11.8^{+9.0}_{-7.0}$ &  15928       \\
\hline
$\Sigma_{Q}^{\ast}\Xi_{QQ}$ & $\frac{3}{2}$ & $1^+$ &  $\dag$                 &  $\dag$       &  $2.8^{+2.5}_{-2.3}$ &  15958 \\
$\Sigma_{Q}^{\ast}\Xi_{QQ}$ & $\frac{3}{2}$ & $2^+$ & $0.1^{+2.0}_{\dag}$     &   ?           &$16.7^{+11.4}_{-9.3}$ &  15944   \\
\hline
$\Sigma_{Q}\Xi_{QQ}^{\ast}$ & $\frac{3}{2}$ & $1^+$ & $3.1^{+6.3}_{-3.0}$ &   6180 & $28.1^{+16.7}_{-14.1}$ &  15936    \\
$\Sigma_{Q}\Xi_{QQ}^{\ast}$ & $\frac{3}{2}$ & $2^+$  & $\dag$ &  $\dag$  & $3.1^{+3.9}_{-2.5}$ &  15961\\
\hline
$\Sigma_{Q}^{\ast}\Xi_{QQ}^{\ast}$ & $\frac{3}{2}$ & $0^+$ & $7.8^{+10.3}_{-6.4}$ &   6237   & $40.9^{+22.2}_{-19.2}$ &   15944 \\
$\Sigma_{Q}^{\ast}\Xi_{QQ}^{\ast}$ & $\frac{3}{2}$ & $1^+$ & $4.2^{+7.4}_{-3.9}$ &   6230    & $30.1^{+17.6}_{-14.9}$ &  15935  \\
\hline
$\Sigma_{Q}^{\ast}\Xi_{QQ}^{\ast}$ & $\frac{3}{2}$ & $2^+$ & $0.3^{+2.3}_{\dag}$ &   ?  & $14.4^{+10.3}_{-8.3}$ &  15971   \\
$\Sigma_{Q}^{\ast}\Xi_{QQ}^{\ast}$ & $\frac{3}{2}$ & $3^+$ & $\dag$ &   $\dag$   & $0.5^{+1.6}_{\dag}$&   $?$  \\
  \hline\hline
\end{tabular}
\end{table}

Unlike the EFT approach,  the isospin-1/2 and isospin-3/2 sectors are correlated in the OBE model. Thus, we can calculate the spectrum of the $\Xi_{cc}^{(\ast)}\Sigma_{c}^{(\ast)}$ system with $I=3/2$, and the results are displayed in Table~ \ref{tab:binding-hidden-dxicc190}.  Unlike the $I=1/2$ case, not all the possible combinations have attraction strong enough to bind. As a matter of fact, only three states, $1^{+}$ $\Sigma_{c}\Xi_{cc}^{\ast}$,  $0^{+}$ $\Sigma_{c}^{\ast}\Xi_{cc}^{\ast}$ and  $1^{+}$ $\Sigma_{c}^{\ast}\Xi_{cc}^{\ast}$  , bind within the uncertainties induced by the breaking of HADS.

Heavy quark flavor symmetry dictates that in our above study, we can replace the charm quark with
its bottom counterpart, and the interactions will remain the same, again up to corrections of $1/m_b-1/m_c$. As a result, we can study the spectrum of the $\Xi_{bb}^{(\ast)}\Sigma_{b}^{(\ast)}$ system with both $I=1/2$ and $I=3/2$. Naively, as the constituents of the system become heavier, the binding energy becomes larger, because the interaction remains the same but the kinetic energy of the system decreases.   This can be clearly seen in Tables \ref{tab:binding-hidden-dxicc189} and \ref{tab:binding-hidden-dxicc190}. All the ten
states in $I=1/2$ bind, but not all of the $I=3/2$ states. For instance, it is quite likely that the $3^+$ $\Sigma_{b}^{\ast}\Xi_{bb}^{\ast}$ state does not bind.

\section{Summary and conclusion}
\label{sec:conclusions}
Motivated by the experimental discovery of $P_c(4312)$, $P_c(4440)$, and $P_c(4457)$ by the LHCb collaboration and
the fact that they could well be part of a complete multiplet of $\bar{D}^{(*)}\Sigma_c^{(*)}$ molecules, we have studied the
$\Xi_{cc}^{(*)}\Sigma_c^{(*)}$ system in the OBE model.
 Ten triply charmed dibaryons with isospin 1/2 are found to  bind, consistent with scenario B of the previous EFT study~\cite{Pan:2019skd}. In addition, ten triply bottom dibaryons are predicted in the isospin-1/2 sector. As for the isospin-3/2 sector, only three molecular states are likely in the charm sector, while nine are possible in the bottom sector.

It should be noted that  the $1^+$ $\Xi_{cc}\Sigma_c$ state has a larger mass than its $0^+$ counterpart in the OBE model, consistent with HADS as predicted in the previous EFT study, which is different from the calculations of Ref.~\cite{Chen:2018pzd} because they kept the delta potential in the OBE model~\cite{Chen:2018pzd}.  We found that by removing the delta potential of the spin-spin component in  the OBE model, we could describe the three LHCb pentaquark states with a single cutoff in the hadronic molecular picture~\cite{Liu:2019zvb}, thus, here we also removed the delta potential in the OBE model.   In Ref.~\cite{Wang:2019gal} Wang employed QCD sum rules to calculate the masses of  $1^+$ and $0^+$ $\Xi_{cc}\Sigma_c$, which, however, cannot determine the relative ordering of these two states because of the relatively large uncertainties  of their approach.  Future lattice QCD studies of these systems
will provide a nontrivial test not only of HADS but also of the molecular nature of the pentaquark states.

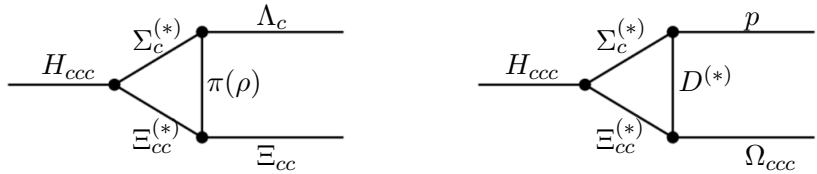
\begin{figure}[!h]
\centering
\begin{overpic}[scale=.6]{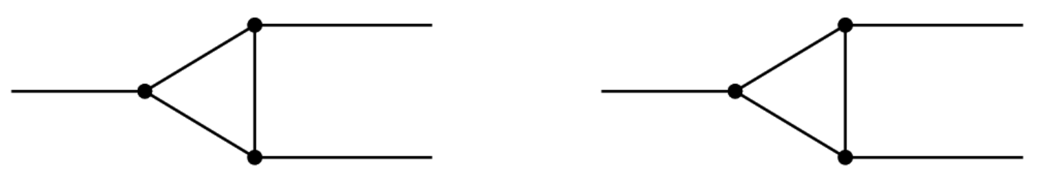}
\put(16,12){$\Sigma_{c}^{(\ast)}$}  \put(72,12){$\Sigma_{c}^{(\ast)}$}
\put(16,0){$\Xi_{cc}^{(\ast)}$}   \put(72,0){$\Xi_{cc}^{(\ast)}$}
\put(25,7){$\pi$$(\rho)$} \put(82,7){$D^{(\ast)}$}
\put(31,15){$\Lambda_{c}$}   \put(90,15){$p$}
\put(31,-2){$\Xi_{cc}$}   \put(90,-2){$\Omega_{ccc}$}
\put(5,9){$H_{ccc}$}  \put(61,9){$H_{ccc}$}
\end{overpic}
\caption{ Likely strong decays of $\Xi_{cc}^{(\ast)}\Sigma_{c}^{(\ast)}$ molecules via triangle diagrams to $\Xi_{cc}\Lambda_{c}$ and $\Omega_{ccc}p$.   \label{xyz}}
\end{figure}

To help experimental searches for the predicted states, it would be of great value to study the decay and productions of these states. Although it is beyond the scope of the present work, it is instructive to point out that some of the states can decay strongly, as shown in
Fig.~\ref{xyz} for $\Sigma_c^{(*)}\Xi_{cc}^{(*)}$ molecules, e.g., via triangle diagrams to $p\Omega_{ccc}$ or $\Lambda_c\Xi_{cc}$. A more quantitative study of such decays will be left for a future work.

\section{Acknowledgments}
 This work is partly supported by the National Natural Science Foundation of China under Grants Nos.11735003, 11975041, and 11961141004, and the fundamental Research Funds
for the Central Universities.

\bibliography{XiccSigmac}

\end{document}